\documentclass[fleqn,12pt,twoside]{article}
\usepackage[headings]{espcrc1}

\readRCS
$Id: espcrc1.tex,v 1.2 2004/02/24 11:22:11 spepping Exp $
\usepackage{graphicx}

\newcommand{\bra}[1]{\langle #1|}
\newcommand{\ket}[1]{|#1 \rangle}
\newcommand{\pb}{{\bf p}}

\newcommand{\ve}{\varepsilon}

\newcommand{\bold}[1]{\mbox{\boldmath $#1$}}
\newcommand{\brn}{\begin{eqnarray*}}
\newcommand{\ern}{\end{eqnarray*}}
\title{Weak nonmesonic decay spectra of hypernuclei}

\author{Eduardo Bauer\address{Facultad de Ciencias Exactas,
           Departamento de F\'{\i}sica,
           Universidad Nacional de La Plata, 
           1900 La Plata, Argentina},
        Alfredo P. Gale\~ao\address{Instituto de F\'{\i}sica Te\'orica,
           UNESP - Universidade Estadual Paulista, \\
           Caixa Postal 70532-2,
           01156-970 S\~ao Paulo, SP, Brazil},
        Mahir S. Hussein\address{Departamento de F\'{\i}sica Matem\'atica,
           Instituto de F\'{\i}sica, Universidade de S\~ao Paulo, \\
           Caixa Postal 66318, 05315-970 S\~ao Paulo, SP, Brazil},
        and
        Franjo Krmpoti\'c\address{Instituto de F\'{\i}sica La Plata, CONICET,
           Universidad Nacional de La Plata, \\
           1900 La Plata, Argentina}\thanks{\textit{E-mail address:}
           krmpotic@fisica.unlp.edu.ar}}

\runtitle{Weak nonmesonic decay spectra of hypernuclei}
\runauthor{E. Bauer, A. P. Gale\~ao, M. S. Hussein and F. Krmpoti\'c}

\begin{document}

\maketitle

\begin{abstract}
We compute one- and two-nucleon kinetic-energy spectra and
opening-angle distributions for the nonmesonic weak decay of
several hypernuclei, and compare our results with some recent
data. The decay dynamics is described by transition potentials of
the one-meson-exchange type, and the nuclear structure aspects by
two versions of the independent-particle shell model (IPSM). In
version IPSM-a, the  hole states are treated as
stationary, while in version IPSM-b the deep-hole ones are
considered to be quasi-stationary and are described by
Breit-Wigner distributions.
\vspace{1pc}
\end{abstract}

The mesonic mode, $\Lambda \to N\pi$, with a rather small $Q$-value,
$Q_{M}=M_\Lambda - M_N -m_\pi \approx 37$ MeV, is heavily
inhibited for $\Lambda$-hypernuclei, except for the very lightest, due to Pauli
blocking.
With increasing mass number, $A$, a new mode quickly becomes dominant,
namely the nonmesonic weak decay (NMWD), $\Lambda N \to nN$, whose $Q$-value,
$Q_{NM} = M_\Lambda - M_N + \varepsilon_\Lambda + \varepsilon_N \approx
120$ -- $135$ MeV, is sufficiently large to render this Pauli blocking less and
less effective. NMWD can be seen as one of the most radical transmutations of an
elementary particle inside the nuclear medium: the strangeness is changed by
$\Delta S=-1$ and the mass by $\Delta M = M_\Lambda - M_N = 176$ Mev. From a
practical point of view, the main interest in NMWD is that it is, at present,
the only way available to probe the strangeness-changing interaction between
baryons.

Lately, the quality of experimental data on NMWD has improved considerably,
and today one has available, not only one-nucleon kinetic energy spectra, but
also two-nucleon coincidence spectra and opening-angle distributions obtained
in several laboratories around the world, such as, KEK, FINUDA, and  BNL.
Here we  briefly discuss a simple but fully quantum-mechanical formalism
for the theoretical investigation of these observables. For
more details see  Refs. \cite{Bar08} and \cite{Bau09}.

We start from Fermi's golden rule for the NMWD transition rate,
$\Gamma_N \equiv \Gamma(\Lambda N \to nN)$,
with the states of the emitted
nucleons approximated by plane waves and initial and final short-range
correlations implemented at a simple Jastrow-like level.
The initial hypernuclear state is taken as a $\Lambda$ in single-particle
state $j_\Lambda=1s_{1/2}$ weakly coupled to an $(A-1)$ nuclear core of spin
$J_C$, i.e., $\ket{J_I}=\ket{(J_Cj_\Lambda)J_I}$.
For the description of nuclear states we adopt the independent particle shell 
model (IPSM). 

Let us consider first the simplest version of this model, IPSM-a, in
which all the relevant particle and hole states are assumed to be stationary.
Thus, if the nucleon inducing the decay is in state $j_N$, then
the possible states of the residual nucleus are
$\ket{J_F}=\ket{(J_Cj_N^{-1})J_F}$ and
the liberated energy is 
$\Delta_{j_N} = M_\Lambda - M + \varepsilon_{j_\Lambda} + \varepsilon_{j_N}$,
where the $\varepsilon$'s are single-particle energies.
Within this scheme, we get 
\begin{eqnarray}
\Gamma_N &=& 2\pi \sum_{Sj_NJ_F}
\int \int
|\bra{\pb_n\pb_N S;J_F}V\ket{J_I}|^2
\delta(E_n+E_N+E_r-\Delta_{j_N})
\frac{d{\bf p}_n}{(2\pi)^3}\frac{d{\bf p}_N}{(2\pi)^3}
\label{1}
\\ 
&=& \frac{4M^3(A-2)}{\pi} \sum_{j_N}
\int dE_N\int dE_n \,\mathcal{F}_{j_N}(p,P),
\label{2}
\end{eqnarray}
where $E_{n,N}=\pb_{n,N}^2/(2M)$ are the kinetic energies of the emitted
nucleons, $E_r=|\pb_n+\pb_N|^2/[2(A-2)M]$ accounts for the recoil, 
and $V$ is the transition potential.
It is understood that all integrations on kinematical variables 
run over the allowed phase space. 
In Eq.~(\ref{2}), 
\begin{eqnarray}
\mathcal{F}_{j_N}(p,P) &=& \sum_{J=|j_N-1/2|}^{j_N+1/2} F_{j_N}^J
\sum_{SlL\lambda J} |\bra{plPL\lambda SJ}V\ket{j_\Lambda j_N J}|^2,
\label{3}
\end{eqnarray}
where $F_{j_N}^J$ are spectroscopic factors,
$p$ and $P$ are the relative and total momenta of the emitted nucleons,
$l$ and $L$ are the corresponding orbital angular momenta, 
and the couplings
$\bold{l}+\bold{L} = \bold{\lambda}$,
$\bold{\lambda}+\bold{S} = \bold{J}$ and
$\bold{j}_\Lambda + \bold{j}_N = \bold{J}$ are performed. The one-nucleon
transition probability density $S_N(E_N)$ is obtained by
taking the derivative of $\Gamma_N$, i. e.,
\begin{eqnarray}
S_N(E_N) = \frac{d\Gamma_N}{dE_N} &=& 
\frac{4M^3(A-2)}{\pi} \sum_{j_N} 
\int dE_n \,\mathcal{F}_{j_N}(p,P). 
\label{4}
\end{eqnarray}
The proton and neutron spectra are then given by
$\Delta N_p(E) \propto S_p(E)$ and
$\Delta N_n(E) \propto S_p(E) + 2S_n(E)$.
Similar developments, with the appropriate choice of kinematical integration
variables, yield the coincidence energy spectra
$\Delta N_{nN}(E) \propto S_{nN}(E)$ and the opening-angle distributions
$\Delta N_{nN}(\cos\theta) \propto S_{nN}(\cos\theta)$, with $N=n,p$.
More details, including the way to fix the normalization of
these spectra and distributions, are given in Ref.~\cite{Bau09}.

For p-shell and heavier hypernuclei, some of the $\ket{j_N^{-1}}$
are deep-hole states, having considerable spreading widths,
$\gamma_{j_N}$, as
revealed for instance in quasifree $(p,2p)$ reactions.
It is, therefore, unreasonable to treat such cases as stationary, zero-width,
states. Rather, they are better approximated as Breit-Wigner distributions in
the liberated energy $\varepsilon$,
$
{\rm P}_{j_N}(\ve)=\frac{2\gamma_{j_N}}{\pi}
\frac{1}{\gamma_{j_N}^2+4(\ve-\Delta_{j_N})^2}.
$
This leads to a slightly more sophisticated version of the IPSM, which we call
IPSM-b. It turns out that the final expressions we need can be obtained from
those of IPSM-a through the replacements:
$
\Delta_{j_N} \mapsto \ve,
$
and
$
\sum_{j_N} \cdots \mapsto
\sum_{j_N} \int_{-\infty}^{+\infty} d\ve {\rm P}_{j_N}(\ve) \cdots .
$
This is explained in more detail, for the case of $S_{nN}(E)$, in
Ref.~\cite{Bar08}.

\begin{figure}[htb]
\begin{minipage}[t]{80mm}
\includegraphics[width=1.0\linewidth]{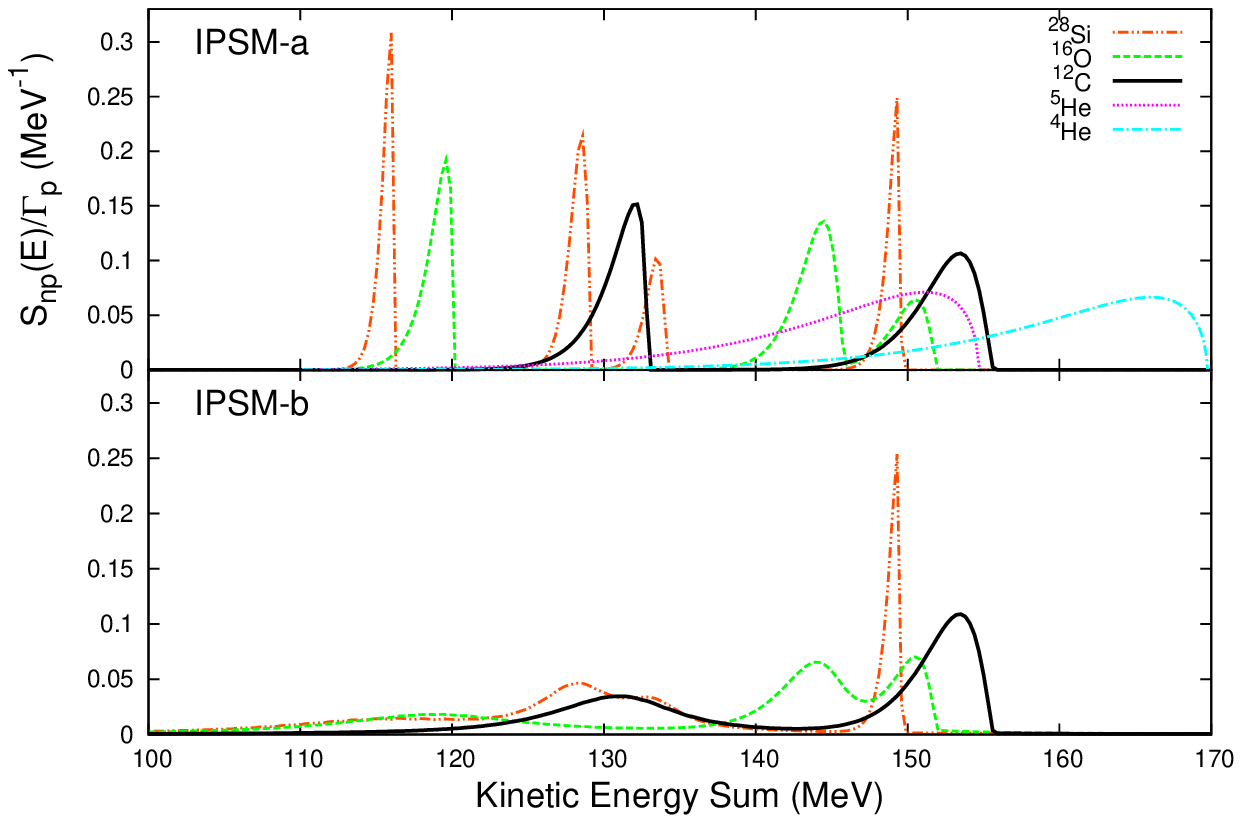}
\caption{\label{Fig1} Normalized energy spectra  $ S_{np}(E)/\Gamma_p$  for
$^{4}_\Lambda$He, $^{5}_\Lambda$He, $^{12}_\Lambda$C, $^{16}_\Lambda$O, and
$^{28}_\Lambda$Si hypernuclei, taken from Ref. \cite{Bar08}. }
\end{minipage}
\hspace{\fill}
\begin{minipage}[t]{75mm}
\includegraphics[width=1.0\linewidth]{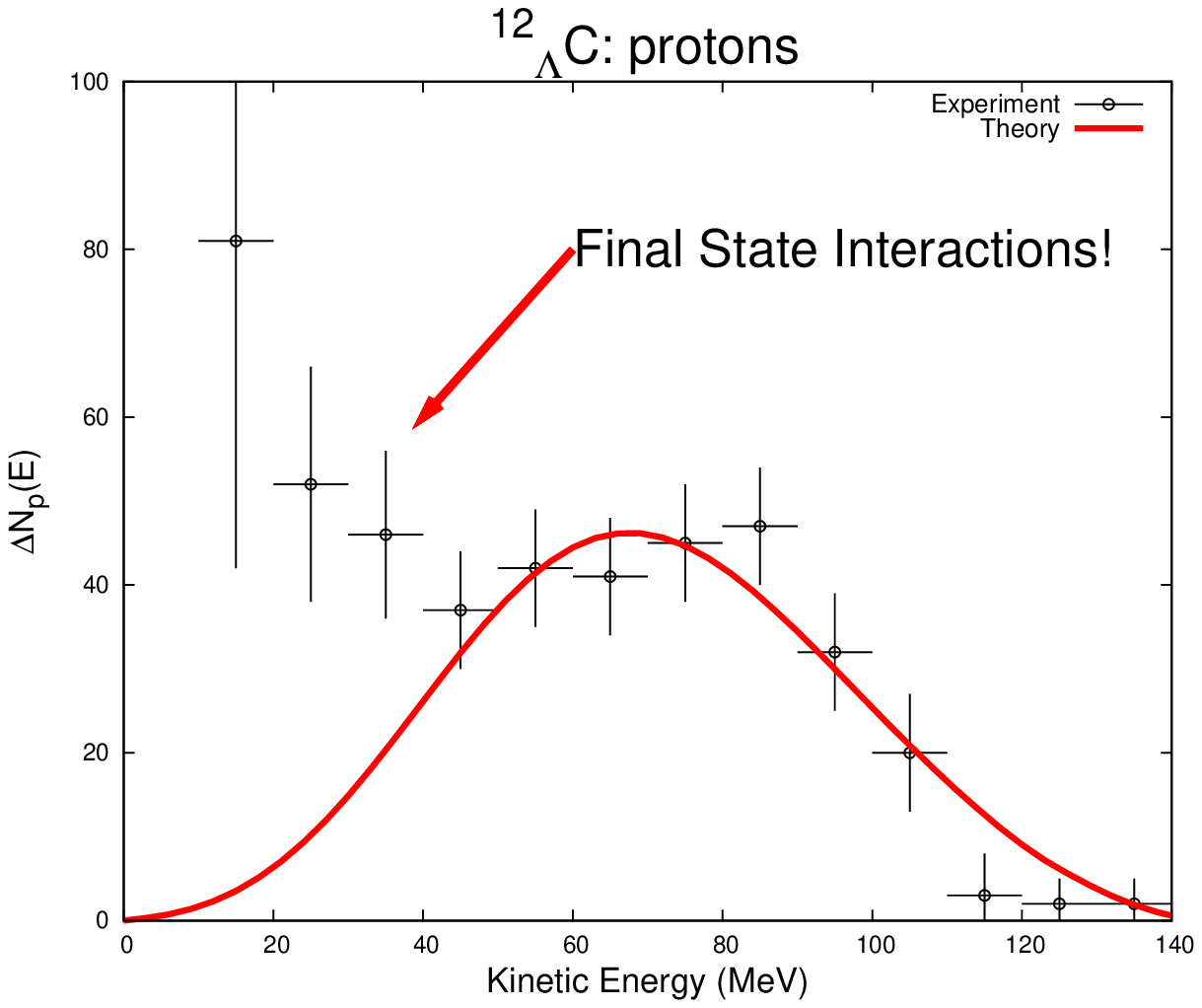}
\caption{\label{Fig2} Comparison between the experimental \cite{Agn08} and
theoretical kinetic energy spectra for protons from $^{12}_\Lambda$C decay. }
\end{minipage}
\end{figure}

The full one-meson exchange potential (OMEP), that comprises the
($\pi,\eta,K,\rho,\omega,K^*$) mesons, has been employed for $V$ in numerical
evaluations.
In Fig. \ref{Fig1} we confront the two IPSM approaches for the spectra
$ S_{np}(E)$.
One sees that, except for the ground states, the narrow peaks engendered by the
recoil effect within the IPSM-a become pretty  wide  bumps within the IPSM-b.

On the other hand, preliminary calculations indicate that the two IPSM 
versions yield similar results  for $S_{N}(E)$ and $S_{nN}(\cos\theta)$.
In Fig. \ref{Fig2} are compared the experimental \cite{Agn08} and theoretical
kinetic energy spectra $\Delta {\rm N}_{p}(E)$
for $^{12}_\Lambda$C, using the just mentioned OMEP.
The theoretical spectrum  is peaked around $85$ MeV, and
reproduces quite well the data for energies larger than  $50$ MeV.
Yet, it differs quite a lot at smaller energies, where the effect
of final state interactions (FSI) is likely to be quite important.

The IPSM reproduces  well \cite{Bau09} the BNL experiment for $^4_\Lambda$He
\cite{Par07}, but it does not reproduce well  the  FINUDA experiment for the
$S_{N}(E)$ spectra in  $^5_\Lambda$He, $^7_\Lambda $Li, and
$^{12}_\Lambda $C \cite{Agn08}.
Once normalized to the transition rate, all the spectra are
tailored basically by the kinematics of the corresponding phase
space, depending very weakly on the dynamics governing the
$\Lambda N \to nN$ transition proper.
The IPSM is the appropriate lowest-order approximation for the
theoretical description of the NMWD of
hypernuclei. It is in comparison to this picture that one should
appraise the effects of the FSI and of the
two-nucleon-induced decay mode.

\end{document}